\documentclass[runningheads,orivec]{llncs}
\usepackage[T1]{fontenc}

\usepackage{graphicx,verbatim}
\usepackage[dvipsnames]{xcolor}
\usepackage{amsmath}

\usepackage{amssymb}
\definecolor{Sblue}{HTML}{1f77b4}
\definecolor{Sorange}{HTML}{bcbd22}
\definecolor{Sgreen}{HTML}{2ca02c}
\definecolor{Sred}{HTML}{d62728}
\definecolor{Spurple}{HTML}{9467bd}

\usepackage{hyperref}       % hyperlinks
\usepackage{xcolor}
\usepackage{sidecap}
\hypersetup{
    colorlinks,
    linkcolor={red!50!black},
    citecolor={blue!50!black},
    urlcolor={blue!50!black}
}

\makeatletter
\newcommand{\printfnsymbol}[1]{%
  \textsuperscript{\@fnsymbol{#1}}%
}
\makeatother

\begin{document}
\title{UNSURF: Uncertainty Quantification for Cortical Surface Reconstruction of Clinical Brain MRIs} 
\titlerunning{UNSURF: Uncertainty Quantification for Cortical Surface Reconstruction}

\author{Raghav Mehta\inst{1,\thanks{Equal contribution}} \and Karthik Gopinath\inst{2,\printfnsymbol{1}} \and Ben Glocker\inst{1,\thanks{Equal senior author}} \and Juan Eugenio Iglesias\inst{2,\printfnsymbol{2}}}  %% Added for anonymized 
\authorrunning{Mehta and Gopinath et al.}
\institute{Imperial College London, UK. \\ \and Athinoula A. Martinos Center for Biomedical Imaging,Massachusetts General Hospital and Harvard Medical School, Boston, USA. \\
    \email{\{raghav.mehta,b.glocker\}@imperial.ac.uk}, \email{\{kgopinath,jiglesiasgonzalez\}@mgh.harvard.edu}}

\maketitle              % typeset the header of the contribution
\begin{abstract}
    We propose UNSURF, a novel uncertainty measure for cortical surface reconstruction of clinical brain MRI scans of any orientation, resolution, and contrast. It relies on the discrepancy between predicted voxel-wise signed distance functions (SDFs) and the actual SDFs of the fitted surfaces. Our experiments on real clinical scans show that traditional uncertainty measures, such as voxel-wise Monte Carlo variance, are not suitable for modeling the uncertainty of surface placement. Our results demonstrate that UNSURF estimates correlate well with the ground truth errors and: \textit{(i)}~enable effective automated quality control of surface reconstructions at the subject-, parcel-, mesh node-level;  and \textit{(ii)}~improve performance on a downstream Alzheimer's disease classification task.

\end{abstract}

\section{Introduction}

    Cortical surface reconstruction from MRI plays a crucial role in neuroimaging, enabling analyses like parcellation and cortical thickness -- a powerful biomarker for studying healthy aging and neurodegenerative diseases such as Alzheimer's~\cite{salat2004thinning}. Traditional methods rely on high-resolution, isotropic MRI scans, typically acquired using T1-weighted (T1) sequences with near 1~mm isotropic resolution. However, clinical MRI scans often exhibit significant variations in orientation, resolution (slice spacing), and contrast, making surface reconstruction difficult.
    
    Recent advancements in machine learning algorithms have enabled  analysis of clinical brain MRI acquired at hospitals \cite{billot2023synthseg,iglesias2023synthsr}. Uncertainty estimation is a critical aspect of cortical surface reconstruction in these settings, since large slice spacing and potential motion artifacts often make surface placement ambiguous. Quantification of uncertainty is required to identify unreliable measurements and control their impact in downstream analyses like group studies of cortical thickness. Despite its importance, uncertainty quantification remains underexplored in cortical surface reconstruction because standard techniques are not applicable due to the topological and geometric constraints involved (e.g., lack of self-intersections).  Without confidence estimates, it is difficult to determine whether surface errors arise from model limitations or poor-quality input data, limiting the clinical reliability and applicability of automated pipelines.
    
    In this work, we introduce UNSURF, a novel measure of uncertainty  in cortical surface reconstructions. UNSURF enables a multi-level Quality Control (QC) strategy at the subject, parcel, and mesh node levels. Experiments on two clinical datasets show that it effectively identifies unreliable regions. We further show that using this strategy for filtering unreliable measurements improves effect sizes in downstream neuroimaging studies of cortical thickness. 
    
    \begin{figure}[t]
        \centering
        \includegraphics[width=1.0\linewidth]{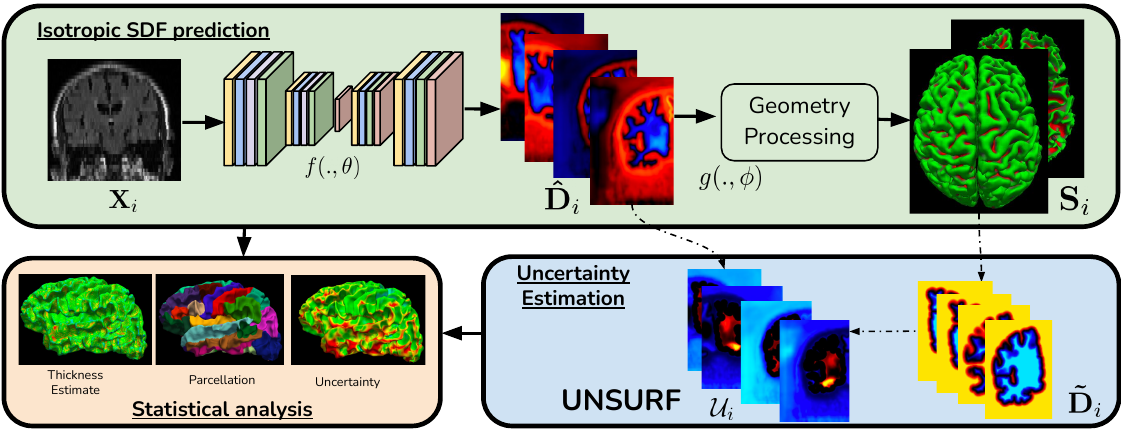}
        \caption{Overview of the proposed uncertainty estimation method (UNSURF).}
        \label{fig:unsurf}
    \end{figure}

    \subsubsection{Related Work:}
        Classical surface reconstruction methods like FreeSurfer~\cite{fischl1999cortical} or BrainSuite~\cite{shattuck2002brainsuite} employ volumetric segmentation followed by  topology correction and geometry-constrained surface extraction. These pipelines enable accurate white and pial surface reconstructions from T1 scans by leveraging neuroanatomical priors. However, they require high-resolution isotropic scans and have long processing times, often exceeding a few hours per subject. Several deep learning methods have been proposed to reconstruct cortical surfaces from MRI scans; they either utilize a combination of convolution and graph-based neural networks~\cite{henschel2022fastsurfervinn,bongratz2022vox2cortex,ma2022cortexode,hoopes2021topofit} or leverage implicit representations~\cite{cruz2021deepcsr,gopinath2021segrecon,ma2022cortexode,zheng2024coupled} to reconstruct cortical surfaces. Nevertheless, these are supervised methods trained for a specific contrast and resolution (isotropic T1 scans) and do not generalize to clinical MRI.
        To address these challenges, a domain randomization approach (\textit{recon-all-clinical}) has recently been introduced~\cite{gopinath2023cortical,gopinath2024synthetic}, which can extract surfaces from scans of any resolution or contrast without domain adaptation or finetuning. This approach builds on prior work on segmentation, synthesis, and super-resolution~\cite{billot2023synthseg,iglesias2023synthsr} and relies on training neural networks on synthetic datasets spanning an unrealistically broad range of (random) contrasts and resolutions, enabling models to generalize across heterogeneous clinical data at test time.
            
        Several methods have been proposed for uncertainty quantification in medical imaging~\cite{band2022benchmarking,huang2024review}. Bayesian deep learning~\cite{BDLneal}, Monte Carlo dropout~\cite{MCD}, and deep ensembles~\cite{DeepEnsemble} have been applied to various medical image analysis tasks~\cite{mehta2021propagating}, such as segmentation~\cite{BUNet,baumgartner2019phiseg,mehta2022qu}, registration~\cite{hu2025hierarchical,dalca2019unsupervised}, precision medicine~\cite{durso2023improving}, or image reconstruction~\cite{Tanno2020UncertaintyMI,tezcan2022sampling}. Test-time augmentation~\cite{wang2019aleatoric} and distribution-aware models~\cite{ren2023uncertainty} further improve reliability by assessing model sensitivity to variations in input data. However,  direct application of these methods to cortical surface reconstruction is not straightforward due to geometric constraints; as such uncertainty quantification for cortical surface reconstruction remains unexplored.

\section{Methodology}
    \label{sec:mainmeth}  
    UNSURF is illustrated in Fig.~\ref{fig:unsurf} and comprises two modules: one for surface placement and another for uncertainty estimation. The surface placement builds on \textit{recon-all-clinical}~\cite{gopinath2024recon} and consists of two sub-modules: first, a convolutional neural network (CNN) that predicts four 1~mm isotropic signed distance functions (SDFs) from an input MRI scan -- two per hemisphere: pial and white matter (WM) surfaces. The second sub-module uses geometry processing to extract topologically constrained surfaces. The uncertainty estimation module quantifies uncertainty by generating distance maps from the predicted surfaces and using them to compute voxel-wise uncertainty measures. These are mapped back to the surfaces to produce the uncertainty estimates on the reconstructed surfaces.

    \subsection{Cortical surface placement via SDF prediction}
        \label{sec:sdf_estimation}  
        For surface placement, our method uses a hybrid algorithm presented in~\cite{gopinath2024recon}, which combines learning-based estimation of SDFs and classical geometry processing and is agnostic to the resolution and contrast of the input scan. Specifically, given an input brain MRI ($\mathbf{X}^v$), we use a CNN to estimate the four SDFs $\mathbf{\hat{D}}_i^v = f(\mathbf{X}^v; \theta)$, which encode the signed distance of each voxel to the four cortical surfaces. Here, $v$ represents a voxel in a volume, and $i \in [1,2,3,4]$ represents four SDFs. To make the CNN agnostic to resolution and contrast, we employ a domain randomization strategy~\cite{billot2023synthseg,iglesias2023synthsr} that generates synthetic data on the fly during training. Crucially, the orientation, resolution, and contrast of these synthetic scans are sampled from uniform distributions at every minibatch, thus forcing the CNN to learn contrast- and resolution-agnostic features. The training data are automatically generated from existing datasets with 1~mm T1s: 3D segmentations (used to generate the synthetic images) are extracted with SAMSEG~\cite{puonti2016fast}, whereas the pial and WM surfaces (used to generate the ground truth SDFs used as targets) are extracted with FreeSurfer~\cite{fischl1999cortical}.
        
        We synthesize training data with the generative model from ~\cite{billot2023synthseg,iglesias2023synthsr}, which  we summarize  here for completeness. Starting from 3D segmentations, random affine and nonlinear spatial transformations are first applied to both the segmentations and SDFs for geometric augmentation. These transformations approximate true SDFs without explicitly deforming surfaces or recomputing distances, offering a computationally efficient solution in practice. Next, it samples a Gaussian mixture model conditioned on the segmentation to generate Gaussian images. The Gaussian parameters (means, variances) are randomly sampled at every minibatch. Non-isotropic acquisitions are simulated by spatial smoothing in order to mimic a combination of orientation, slice spacing, and thickness -- also sampled randomly at every minibatch. Finally, a (randomly sampled) smooth multiplicative bias field is applied to the images. We note that the voxel size of the input image and the target SDFs is 1~mm isotropic; the \emph{intrinsic} resolution of the input depends on the simulated resolution (and is generally lower than 1~mm isotropic), whereas the SDF is always crisp. This strategy enables the CNN to predict high-resolution SDFs independently of the resolution of the input.  
        
        The neural network is trained to predict the SDFs of the WM and pial surfaces using an L2 loss: $L_{\text{SDF}} = \frac{1}{|v|}\sum_{i} (\mathbf{\hat{D}}_i^v - \mathbf{D}_i^v)^2$, where $\mathbf{D}_i^v$ is the ground-truth SDF. We truncate the SDFs at $\pm5$~mm to constrain learning to cortical regions (e.g., we do not want to waste network capacity by discriminating whether a voxel is 9 vs 10~mm away from the surface). At test time, the input is resampled to 1~mm resolution and  the trained model predicts 1~mm isotropic SDFs for real MRI scans irrespective of their  resolution and contrast, as explained above. 
        
        The predicted distance maps $\mathbf{\hat{D}}_i^v$ are used for surface extraction ($\mathbf{S}_i^n$) using a geometry processing module $g(.,\phi)$ similar to that of~\cite{gopinath2024recon} (which in turn relies heavily on FreeSurfer tools~\cite{fischl1999cortical}). In short, the WM surface $\mathbf{S}_i^{\text{wm}}$ is initialized as the level-set of the predicted WM SDF and is iteratively smoothed and deformed to optimize mesh regularity. Topological correction is performed on the mesh to ensure that its Euler characteristic is 2 (i.e., topologically equivalent to a sphere, wihtout holes or handles). The refined mesh is then inflated by expanding the cortical surface into a sphere while maintaining the anatomical features. The pial surface $\mathbf{S}_i^{\text{pial}}$ is then placed by expanding $\mathbf{S}_i^{\text{wm}}$ outward using the pial SDF while preventing self-intersections. The surfaces are spherically registered to a common coordinate system, enabling cortical parcellation using the Desikan-Killiany Atlas. This step allows subject-wise comparisons and statistical analysis of cortical morphology across populations.

    \subsection{Uncertainty Estimation for Cortical Surface Reconstruction}
    
        \subsubsection{Baseline - Ensemble Dropout:}
            To estimate uncertainty for cortical surface reconstruction, we utilize ensemble dropout~\cite{smith2018understanding} which has been shown to provide better uncertainty estimates for different tasks~\cite{mehta2021propagating,ashukha2020pitfalls,fuchs2021practical} compared to Monte Carlo dropout~\cite{MCD} or deep ensembles~\cite{DeepEnsemble}. Using $N$ independently trained networks, and taking $Z$ dropout samples from each of them, ensemble dropout produces a total $M=N\times Z$ samples of SDFs for the same input.
            The final SDF estimate $\mathbf{\hat{D}}_i^v$ is computed as: $\mathbf{\hat{D}}_i^v = \frac{1}{M} \sum_{m=1}^{M} \mathbf{\hat{D}}_{i_{m}}^v$. The voxel-wise uncertainty is measured as the variance across predictions:
            $$({\sigma^v_i})^2 = \frac{1}{M-1} \sum_{m=1}^{M} (\mathbf{\hat{D}}_{i_{m}}^v - \mathbf{\hat{D}}_i^v)^2.$$
            
            While the variance is useful for capturing local deviations in $\mathbf{\hat{D}}_i^v$, it does not account for topological corrections introduced by $g(\cdot,\phi)$. An alternative approach would be to estimate variance uncertainty on the surface mesh node coordinates $\mathbf{S}_i^n$. However, this would be computationally costly, as each geometry processing step $g(.,\phi)$ takes approximately 2 hours; as such, passing M distance maps through it would require M$\times$2 hours, which is infeasible in a practical scenario. Another approach could be to use surface deformation methods~\cite{bongratz2022vox2cortex,hoopes2021topofit,ma2022cortexode} that do not require explicit topological correction, allowing estimation of variance directly on the surface nodes without high computational overhead. However, to our knowledge, these methods are currently limited to high-resolution T1 scans, and Recon-all-clinical~\cite{gopinath2024recon} is currently the only approach applicable to surface reconstruction of heterogeneous clinical scans. \\

        \subsubsection{UNSURF - Uncertainty estimation with difference map:}
            To overcome variance limitations (high computation time and lack of topological correction), we estimate uncertainty using a new measure (UNSURF), which relies on the consistency (or lack thereof) between the SDFs predicted by the neural network ($\mathbf{\hat{D}}_i^v$) and actual SDFs ($\mathbf{\tilde{D}}_i^v$) computed from the extracted surfaces $\mathbf{S}_i^n$ after geometry processing ($g(\cdot,\phi)$). Voxel-level uncertainty as the squared discrepancy:
            $$\mathcal{U}_i^v = (\mathbf{\hat{D}}_i^v-\mathbf{\tilde{D}}_i^v)^2.$$
            $\mathcal{U}$ serves as a more accurate proxy for uncertainty in surface reconstruction, reflecting geometric corrections and areas with significant discrepancies between the initial voxel-wise prediction ($\mathbf{\hat{D}_i^v}$) and the actual SDF of the final surface ($\mathbf{\tilde{D}_i^v}$). Larger $\mathcal{U}$ indicate greater uncertainty, as they correspond to regions with 
            multiple plausible surfaces.

    \subsection{Implementation details}
        Our voxel-wise regression CNN follows the same architecture as \textit{recon-all-clinical} ~\cite{gopinath2024recon}, utilizing a 3D U-Net~\cite{ronneberger2015u} trained with synthetic pairs generated on the fly, as described in Section~\ref{sec:sdf_estimation}. The U-Net consists of five levels with two layers each, employing $3 \times 3 \times 3$ convolutions and exponential linear unit activations. Each level $l$ contains $24\times2^{(l-1)}$ features. The final layer includes a dropout layer and uses a linear activation function to model SDFs.  
        
        To estimate uncertainty using ensemble dropout variance ($\sigma^2$), we train five randomly initialized models on the same dataset using identical hyperparameters and total epochs. We perform 20 forward passes per input for each of the five models at test time, averaging the resulting SDF predictions ($\mathbf{\hat{D}_i^v}$). The geometry processing pipeline remains the same as that of FreeSurfer. Once the surfaces are reconstructed, we estimate the implicit SDF ($\mathbf{\tilde{D}_i^v}$) to compute the $\mathcal{U}^v_i$. 
        
        We resampled the uncertainty ($\mathcal{U}^v$ and $(\sigma^2)^v$) values onto the mesh nodes using trilinear interpolation,  to obtain node-level uncertainty ($\mathcal{U}^n$ and $(\sigma^2)^n$). Given a parcellation, we compute region-wise uncertainty by averaging it over the nodes in each parcel ($\mathcal{U}^p$ and $(\sigma^2)^p$). Similarly, to obtain a subject-level representation, we compute the global average of uncertainty over all surface nodes ($\mathcal{U}^s$ and $(\sigma^2)^s$). This approach allows us to summarize uncertainty at different levels of granularity, capturing localized effects at the node level and parcel level while also obtaining an overall subject-wise measure.

\section{Experiments and Results}

    \subsection{Datasets}
    
        \textbf{Training Data}:
        We trained the CNN using segmentation maps and SDFs derived from 1,000 1~mm T1 MRI scans. These include 500 scans from the Human Connectome Project (HCP)~\cite{glasser2013minimal} and 500 scans from the Alzheimer's Disease Neuroimaging Initiative (ADNI)~\cite{jack2008alzheimer}.  FreeSurfer~\cite{fischl1999cortical} was used to obtain ground-truth cortical surfaces and morphometric measures, such as cortical parcellation and cortical thickness. We emphasize that only the 3D segmentations~\cite{puonti2016fast} and estimated SDFs from the white and pial surfaces are used during training;  the original T1-weighted MRI scans are disregarded. 
        
        \noindent \textbf{Test Datasets}:
        We used two different datasets to evaluate our method. The first dataset  (henceforth ``synthetically downsampled dataset'') comprises subjects with high-quality scans, which we progressively downsample to analyze the performance of our method as a function of resolution. Specifically, we used 15 subjects chosen at random from the HCP dataset (each with 0.7~mm isotropic T1 and T2 scans) and 15 subjects from the ADNI3 dataset (each with $\sim$1~mm isotropic T1 and FLAIR scans). We simulated clinical acquisition on three different orientations (axial, coronal, sagittal) for 5 different slice spacings (2~mm, 3~mm, 4~mm, 5~mm, and 6~mm). The slice thickness was made equal to the slice spacing in all cases. The second dataset comprises 200 randomly selected subjects (age: 74.0$\pm$7.4 years; 95 males) from ADNI1. Since ADNI1 includes high-resolution ($\sim$1~mm isotropic) T1 scans and a 5~mm axial FLAIRs, it enables evaluation on images natively acquired at low-resolution (the axial FLAIRs) using ground truth derived from the high-resolution T1s.

    \subsection{Comparison Between Variance and $\mathcal{U}$}  
    
        \begin{figure}[t]
            \centering
            \includegraphics[width=1.0\linewidth]{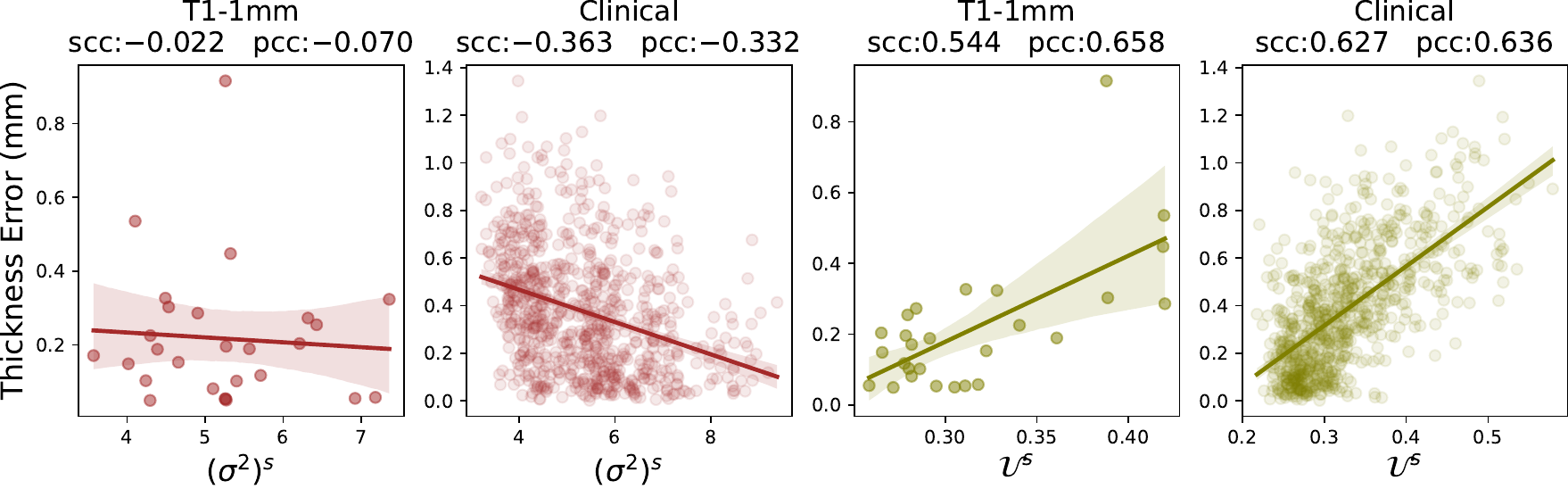}
            \caption{Comparison of uncertainty estimates vs. thickness error for Variance - $(\sigma^2)^s$ (left) and $\mathcal{U}^s$ (right) on high-resolution (T1-1mm) and clinical scans. SCC and PCC stand for Spearman and Person correlation coefficient.}
            \label{fig:GAU_Variance_Comparison}
        \end{figure}  
    
        We first compare \textbf{UNSURF} $\mathcal{U}$ with \textbf{ensemble dropout variance} by analyzing their correlation with \textbf{thickness errors}. Fig.~\ref{fig:GAU_Variance_Comparison} shows the correlation coefficients (Spearmans - SCC, and Pearsons - PCC) between uncertainty estimates and thickness errors at subject level ($\mathcal{U}^s$ and $(\sigma^2)^s$). Being unaware of geometry, variance-based uncertainty exhibits weak or even negative correlation with thickness errors across both high-resolution and clinical-quality scans, making it unreliable for identifying surface reconstruction errors. In contrast, $\mathcal{U}$ consistently shows a strong positive correlation with thickness error, indicating that it better captures regions where cortical surface placement is uncertain. The key reason behind $\mathcal{U}$’s improved performance is that variance only reflects prediction dispersion, whereas $\mathcal{U}$ explicitly accounts for the geometric corrections applied during surface reconstruction. This makes $\mathcal{U}$ a more precise measure of uncertainty in cortical thickness estimation.
    
    \subsection{Results on synthetically downsampled dataset}  
    
        \begin{figure}[t]
            \centering
            \includegraphics[width=1.0\linewidth]{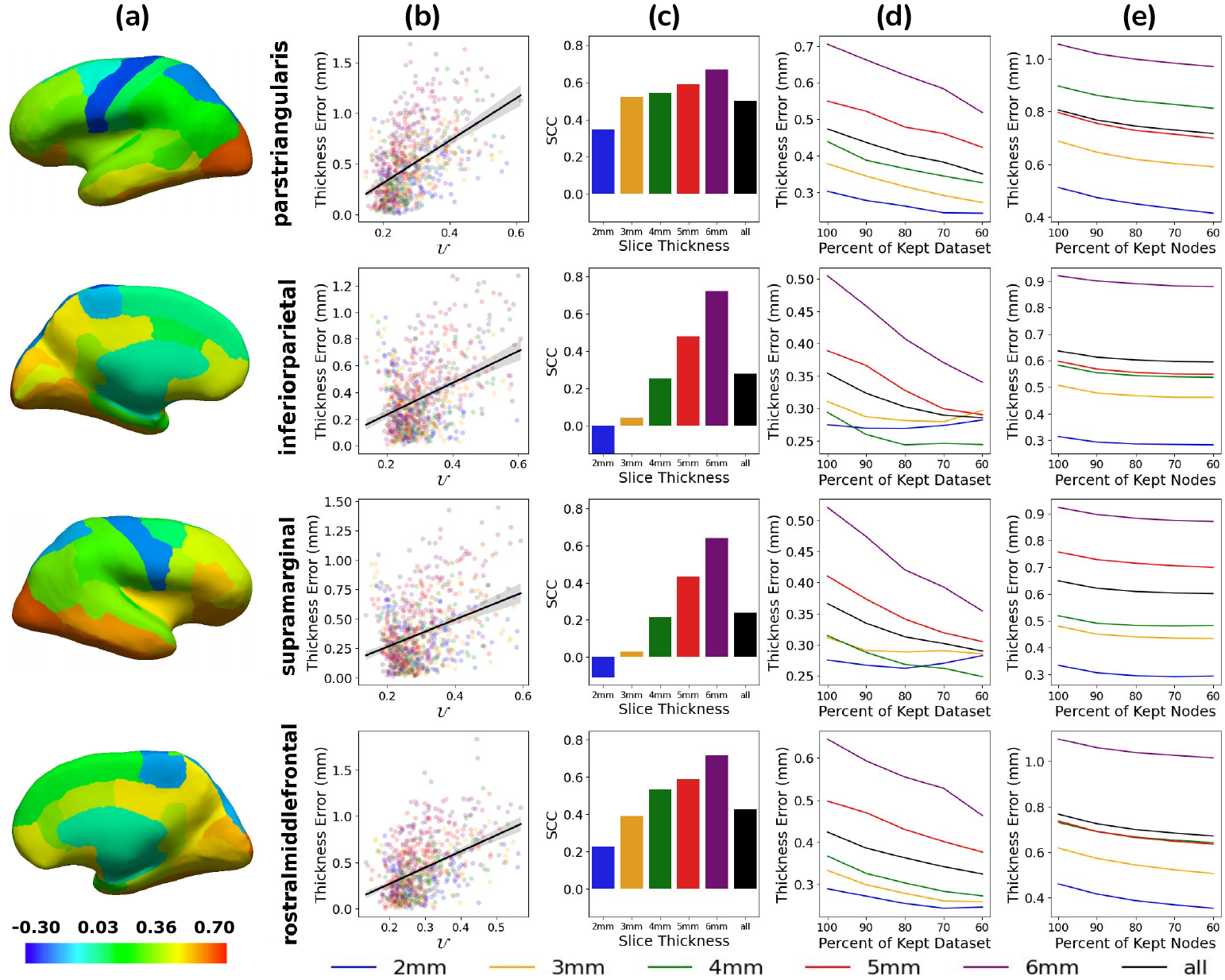}
            \caption{Evaluation of parcel-level and node-level uncertainty estimates on clinical scans. (a)~SCC between uncertainty and thickness error for all parcels. (b)~$\mathcal{U}^p$ vs. thickness error for four different parcels and varying slice thickness. (c)~SCC across slice thicknesses. (d)~Thickness error as a function of percent of kept data based on parcel-level uncertainty ($\mathcal{U}^p$). The x-axis refers to the percent of kept dataset based on uncertainty filtering (i.e. At 100, all subjects are kept when computing thickness error). (e)~Thickness error as a function of percent of kept data based on node-level uncertainty ($\mathcal{U}^n$) within the top-10\% uncertain subjects ($\mathcal{U}^p$).}
            \label{fig:parcel-evaluation}
        \end{figure}  
        
        To better understand how $\mathcal{U}$ behaves across different cortical regions, we show in Fig.~\ref{fig:parcel-evaluation}(a) the spatial distribution of SCC across the cortex, confirming a strong correlation between $\mathcal{U}^p$ and thickness error in many regions. Furthermore, we analyze four representative parcels that are highly relevant to Alzheimer's disease~\cite{salat2004thinning} (parstriangularis, inferiorparietal, supramarginal, and rostralmiddlefrontal). In Fig.~\ref{fig:parcel-evaluation}(b), we observe a clear positive correlation between $\mathcal{U}$ and thickness error for all four selected parcels, further supporting that $\mathcal{U}$ effectively captures reconstruction uncertainty.  Interestingly, we see in Fig.~\ref{fig:parcel-evaluation}(c) that SCC values are lower for thinner slices (e.g., 2~mm and 3~mm), i.e., it is harder to accurately predict differences between smaller errors. 
        
        We further quantify uncertainty by filtering out the most uncertain predictions and examining the change in thickness errors. In Fig.~\ref{fig:parcel-evaluation}(d), we observe that removing high-$\mathcal{U}^p$ parcels reduces overall parcel-level thickness errors, demonstrating that $\mathcal{U}$ can be used for automated QC. Further, in Fig.~\ref{fig:parcel-evaluation}(e), we see that filtering uncertain ($\mathcal{U}^n$) nodes within already uncertain parcels ($\mathcal{U}^p$) leads to further reduction in thickness errors, reinforcing the utility of multi-level QC.  
    
    \subsection{Results on the ADNI Dataset (AD vs. CN Classification)}  
    
        \begin{figure}[t]
            \centering
            \includegraphics[width=1.0\linewidth]{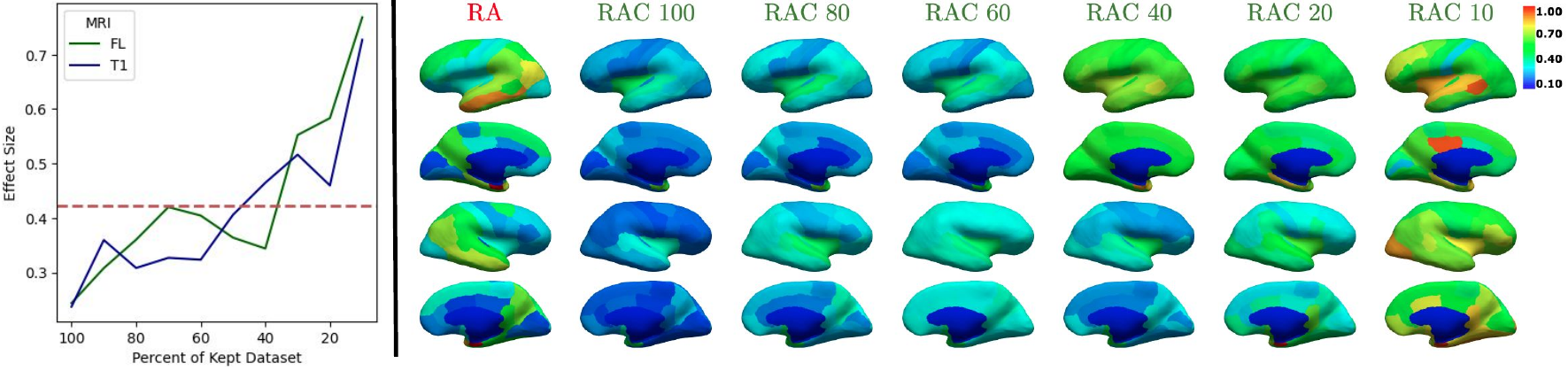}
            \caption{Effect size (ES) as a function of the kept dataset percentage based on uncertainty ($\mathcal{U}^s$). (Left) Mean ES across all parcels for RAC (both FL and T1 MRI). The red dashed line represents ES for Recon-All. (Right) ES per parcel for RA and RAC.}
            \label{fig:effect-size}
        \end{figure}  
        
        We evaluate the impact of $\mathcal{U}$-based QC on a AD vs CN classification task, which is representative of group studies in neuroimaging. The goal is to assess whether removing uncertain cortical thickness measurements ($\mathcal{U}^s$) improves statistical power when measuring group differences. We first fit a general linear model for cortical thickness at each parcel, adjusting for age and gender. We then compute Cohen’s d as effect size  between the AD and CN groups. As shown in Fig.~\ref{fig:effect-size}(left), the effect size increases as we remove high-uncertainty data, indicating that uncertainty-based filtering enhances group separation. The trend of the parcel-wise effect size in Fig.~\ref{fig:effect-size}(right) confirms that removing uncertain thickness estimates improves sensitivity across different brain regions.

\section{Conclusion}
    This paper proposed a novel uncertainty estimation measure (UNSURF - $\mathcal{U}$) for cortical parcel reconstruction using implicit surface methods, and demonstrates its importance in clinical applications. Future work will seek to incorporate uncertainty estimation into faster explicit surface reconstruction methods based on deforming templates~\cite{bongratz2022vox2cortex,hoopes2021topofit,ma2022cortexode}.
    By incorporating uncertainty, our approach improves the robustness and downstream performance of cortical surface analysis on clinical MRI, facilitating large-scale neuroimaging studies on uncurated, heterogeneous clinical datasets. 
        
\section*{Acknowledgment}  
R.M. is funded through the European Union’s Horizon Europe research and innovation programme under grant agreement 10108030. B.G. acknowledges support from the Royal Academy of Engineering as part of his Kheiron Medical Technologies/RAEng Research Chair in Safe Deployment of Medical Imaging AI. K.G and J.E.I are primarily funded by the National Institute of Aging (1R01AG070988, 1RF1AG080371 and 1R21NS138995). Further support is provided by, BRAIN Initiative (1RF1MH123195, 1UM1MH130981), National Institute of Biomedical Imaging and Bioengineering (1R01EB031114).

\subsection*{Disclosure of interests}
B.G. is part-time employee of DeepHealth. No other competing interests.

% \newpage
\bibliography{References}

\end{document}